\documentclass[11pt]{article}
\usepackage{blois,epsfig}

\begin{document}

\footnote{O.I.PISKOUNOVA, e-mail: piskoun@sci.lebedev.ru, tel. 8(499)1326317, P.N.Lebedev Physics Institute of Russian Academy of Science, Leninski pr. 53, 119991 Moscow, Russia}

\vspace{4mm}
\title{The Difference Between Hyperon Transverse Momentum Distributions in $pp$ and $p\bar{p}$ Collisions}

\author{O.I.PISKOUNOVA}

\address{P.N.Lebedev Physics Institute of Russian Academy of Science, Moscow, Russia}

\maketitle\abstracts{The analysis of data on hyperon transverse momentum distributions, $dN/dp_t$, that were gathered from various experiments (WA89, ISR, STAR, UA1, UA5 and CDF) reveals an important difference in the dynamics of multiparticle production in proton-proton vs. antiproton-proton collisions in the region of transverse momenta 0.3 GeV/c$ < p_t < $3 GeV/c. Hyperons produced with proton beam display a sharp exponential slope at low $p_t$, while those produced with antiproton beam do not. Since LHC experiments have proton projectiles, the spectra of multiparticle production at LHC should seem “softer” in comparison to the expectations, because the MC predictions were based on Tevatron (antiproton-proton) data. From the point of view of the Quark-Gluon String Model, the most important contribution to particle production spectra in antiproton-proton reactions is due to antidiquark-diquark string fragmentation that is very interesting object for the futher investigations. This study may have impact not only on
the interpretation of LHC results, but also on cosmic ray physics and astrophysics, where baryon contribution into matter-antimatter asymmetry is being studied.}

\section{Introduction}

The available data of many high energy experiments on $p\bar{p}$ collisions \cite{ua1,cdf} as well as on $pp$ collisions of lower energies \cite{star,isr,wa89} are considered in this article in order to understand the influence of quark composition of beam particles on the shape of transverse momentum spectra of $\Lambda^0$ hyperon production at high energy collider experiments (see Fig.1). Unfortunately, the difference in spectra was not studied enough at ISR energies, where both projectiles were available. 

Obviously, the form of the spectrum at low $p_t$ has strong impact on the value of cross section that is resulting from the integral of transverse momentum distribution over $p_t$.
As it is seen in the Fig.1, the slopes of spectra are changed from lower energy data of proton-proton collision experiment to high energy antiproton-proton S$p\bar{p}$S and Tevatron collider data. What expectations shall we have for proton-proton LHC collider? Here we try to build a concept of
the hyperon spectra for high energy p-p collisions at LHC as well as to explain the difference between Tevatron and LHC data.  

\section{Quark Gluon String model}

The difference in $p_t$ spectra of $\Lambda^0$'s produced in high energy $pp$ and $p\bar{p}$ collisions can not be explained in the QCD theoretical models, because at the collider energies both interactions should give the mutiparticle production due to Pomeron (or multiPomeron) exchange. 

The total cross section and the spectra in $pp$  and $p\bar{p}$ collisions are to be similar because it depends only on the parameters of the Pomeron exchange between two interacting hadrons and should not be sensitive to the quark contents of colliding beams at high energies. So, the difference in the spectra seems a local effect that has no influence on the integral value of cross sections. 

Let us look at pomeron diagrams of proton-proton and antiproton-proton collisions that is shown in the Fig.2.
From the point of view of Quark-Gluon String Model \cite{qgsm,hyperon}, which is based on the Regge theory and on the Pomeron phenomenology, the spectra of produced particles are the results of the cut of pomeron diagram corresponding to the considered reaction. As it seen from the comparison of interaction diagrams, the most important contribution to hyperon production spectra in antiproton-proton reaction is brought by antidiquark-diquark string fragmentation since this part of cylinder configuration certainly takes the most part of energy of colliding $p\bar{p}$. Othervise, the proton-proton collision diagram is symmetric and built from two similar quark-diquark strings.
We are also making here a suggestion that the highest  $p_t$s
in the spectra go from the quark (diquark) of projectiles at the end of the strings.

\section{Comparison of transverse momentum spectra}

The important fact is that the latest experiments of highest energies were
carried out with antiproton beams. It seems the mistake to suggest that $pp$ and $p\bar{p}$ at
high energy are to give the similar hadron transverse momentum distributions. The spectrum of hyperons that was produced with proton beam at STAR \cite{star} has the sharp exponential slope at low $p_t$, while the spectra with antiproton beam are of more flat configuration. The highest energy experiments (UA1 and CDF) show the harder transverse momentum spectra which can not be the result of 
growing energy - it is the result of different form of transverse momentum distributions in different reactions.
 
In the same time, the low $p_t$ sharp exponential contribution seams existing in hyperon spectra in $p\bar{p}$ reaction if we pay attention to the spectrum at $\sqrt{s}$ = 546 GeV in UA5 collaboration \cite{ua5}.  Due to the data at $\sqrt{s}$ = 546 GeV we can conclude that the sharp exponential slope is still there at very low $p_t < $0.7 GeV/c. It gives the hope that this exponential addition exists in other antiproton spectra as well, but it doesn't seen because of the absence of measurements at very low $p_t$.
In Fig.4 the transverse momentum distributions at the collider energies $\sqrt{s}$ = 200 GeV for two experiments, UA5($p\bar{p}$) and STAR ($pp$) are shown.

So, in common, the difference in production cross sections at $pp$ and $p\bar{p}$ collisions seems giving no impact on cross sections. Our study is practically concentrated on the contribution of antidiquark-diquark string fragmentation in antiproton-proton collisions that results in a visible excess in the hadron spectra at the $p_t$ region, 0.8 GeV/c $< p_t < $3. GeV/c, for $p\bar{p}$ reaction. 

The nature of the excess goes from the asymmetric share of energy between the sides of pomeron cylinder and have to be studied in future proton-antiproton experiments of low energy\cite{tapas}. At low energy the annihilation between quark-antiquark takes place and, as it seen from the Fig.5, the resulting spectra consists of only the contribution from diquark-antidiquark string.

\section{LHC expectations and results}

Since LHC experiments have proton projectiles, the spectrum of hyperon production seems "softer" in the comparison to predicted one, because the MC distributions were tuned to the Tevatron data.

The energy splitting in the case of symmetric proton-proton pomeronic diagram should lead, on my openion, to the sharp exponential $p_t$ distribution that goes from very low $p_t$'s up to $p_t^{max}$/2 as it is shown in the Fig.6. In the same time the asymmetric share of energy in $p\bar{p}$ makes
the visible flattering in the spectra , see Fig.7, due to contribution of diquark-antidiquark chain that goes from 1 GeV/c up to 2$p_t^{max}$/3.
All these reasons would be illustrated very well, if we have realy high energy deposit into
transverse momenta, but it seems to me that only fragmentation tail of the spectra was still seen in the existing experiments due to the short strings in $p_t$.

Since the significant amount of data on hyperon distributions is already collected in LHC experiments,
the next paper about this sort of analysis is to be devoted to the detailed comparison of spectra
from ATLAS, ALICE, CMS and LHCb.

\section{Summary} 

The transverse momentum baryon spectra in proton-antiproton collisions (UA1,UA5,CDF) differ 
from the $p_t$ distributions of baryons in proton-proton collisions (STAR, ISR, etc.).

Quark-Gluon String Model explains this phenomenon as a difference in the splittings 
of transverse energy between two sides of pomeron multiparticle production diagram for
one considered reaction vs. another. The diagram for proton-antiproton case includes the unusual string with the diquark-antidiquark ends, which certainly accumulates more energy than another quark-antiquark side of cylinder. It leads to specific contribution that is
boosted to higher $p_t$'s. In this case the hyperon distribution looks like as the spectra in proton-proton collision, but of the energy that is approximately by 30 $\%$ higher.
 Such reason can give the explanation of the form of spectra in
$p\bar{p}$ collision that is systematically harder than the spectra in $pp$ reaction \cite{cosmicrays}.

It is possible as well to suggest that the difference in spectra will disappear with the growing of energy due to the growing multiPomeron contributions into the differencial cross section that are similar as for
$p\bar{p}$ as for $pp$ collisions.

The region of low $p_t$, $p_t < $1 Gev/c, is also interesting for our research because 
the features of diquark-antidiquark string fragmentation has to be seen there too.
This region can be studied in measurements at low energy $p\bar{p}$ collisions.

In order to confirm the theoretical suggestions that are discribed above, I am looking 
forward to analyse the measurements of hyperon transverse momentum spectra from all LHC experiments.
The detail discussion of the results from LHC experiments is going to be done in the next publication.

\section{Acknowledgments}

This article is devoted to the memory of Prof. Alexei Kaidalov, who was my teacher and co-athor
during many years.


\begin{figure}[thb]
\begin{center}
\epsfig{figure=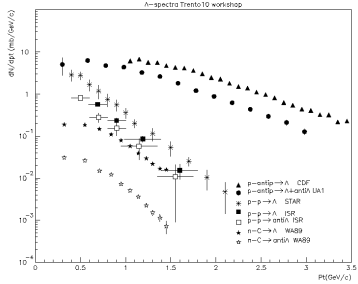,height=3.0in}
\caption{Transverse momentum distributions from different experiments and of various energies:
CDF(1.8TeV), UA1(630GeV), STAR(200GeV), ISR (53GeV) and fixed target exp. WA89.}
\label{fig:1}
\end{center}
\end{figure}


\begin{figure}[thb]
\begin{center}
\epsfig{figure=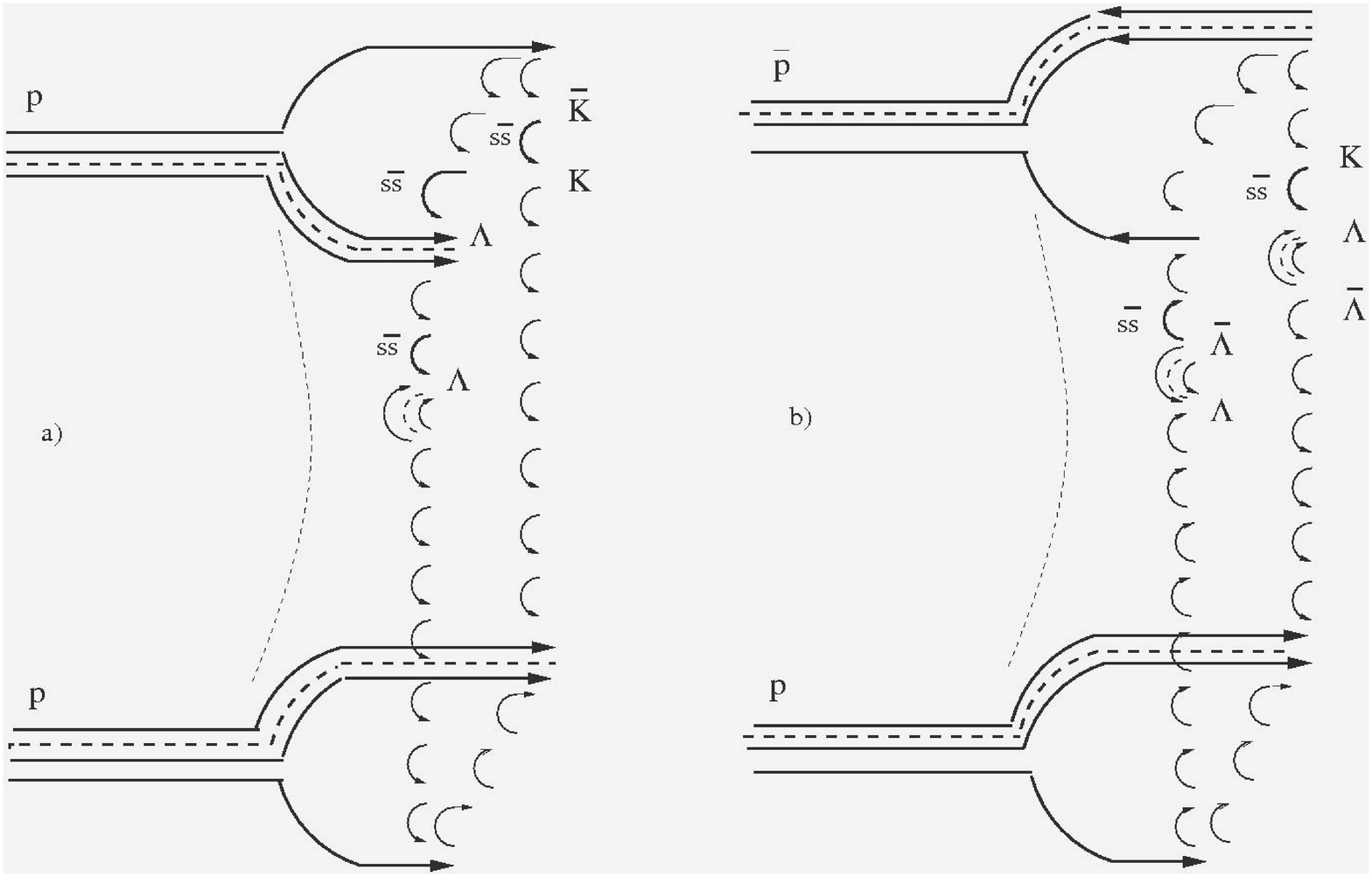,height=2.0in}
\caption{Multiparticle production diagrams for a) $pp$ and b)$p\bar{p}$ reactions.}
\label{fig:2}
\end{center}
\end{figure}


\begin{figure}[thb]
\begin{center}
\epsfig{figure=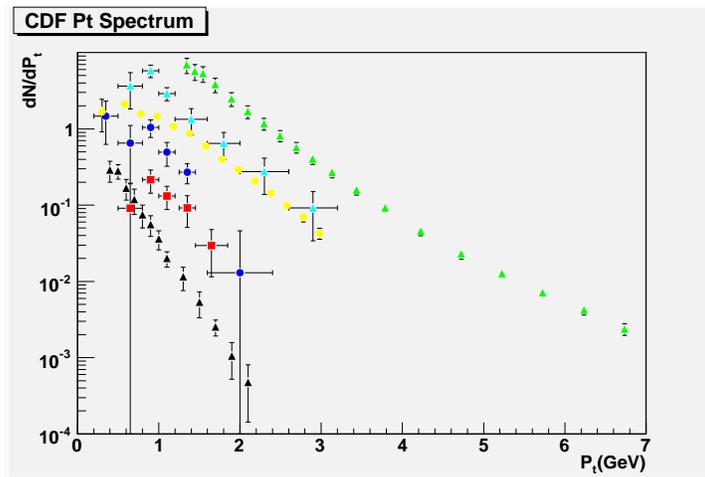,height=2.5in}
\caption{Transverse momentum distributions from different experiments and of various energies:
CDF(1800GeV)-green triangles, UA5(900GeV)-aqua triangles, UA1(630GeV)-yellow circles, UA5(546GeV)
-blue circles, UA5(200GeV)-red squares, STAR(200GeV)-black triangles.}
\label{fig:3}
\end{center}
\end{figure}


\begin{figure}[thb]
\begin{center}
\epsfig{figure=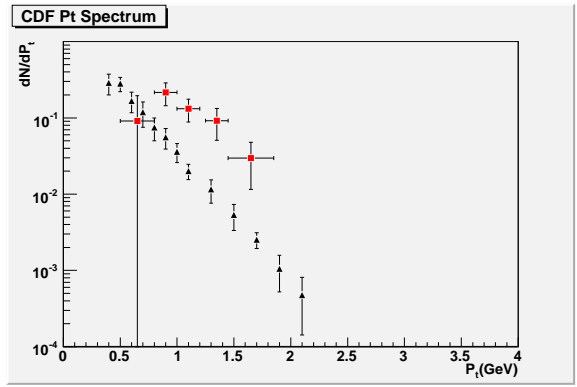,height=2.0in}
\caption{Transverse momentum distributions from the experiments at $\sqrt(s)$ = 200 GeV:
UA5($p\bar{p}$)-red squares and STAR($pp$)-black triangles.}
\label{fig:4}
\end{center}
\end{figure}


\begin{figure}[thb]
\begin{center}
\epsfig{figure=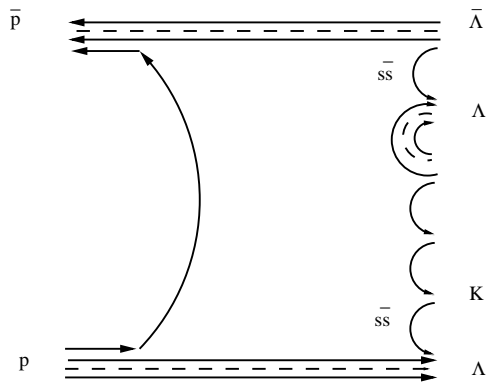,height=2.0in}
\caption{Multiparticle production diagram for $p\bar{p}$ reaction with $\sqrt{s}<$10 GeV.}
\label{fig:5}
\end{center}
\end{figure}


\begin{figure}[thb]
\begin{center}
\epsfig{figure=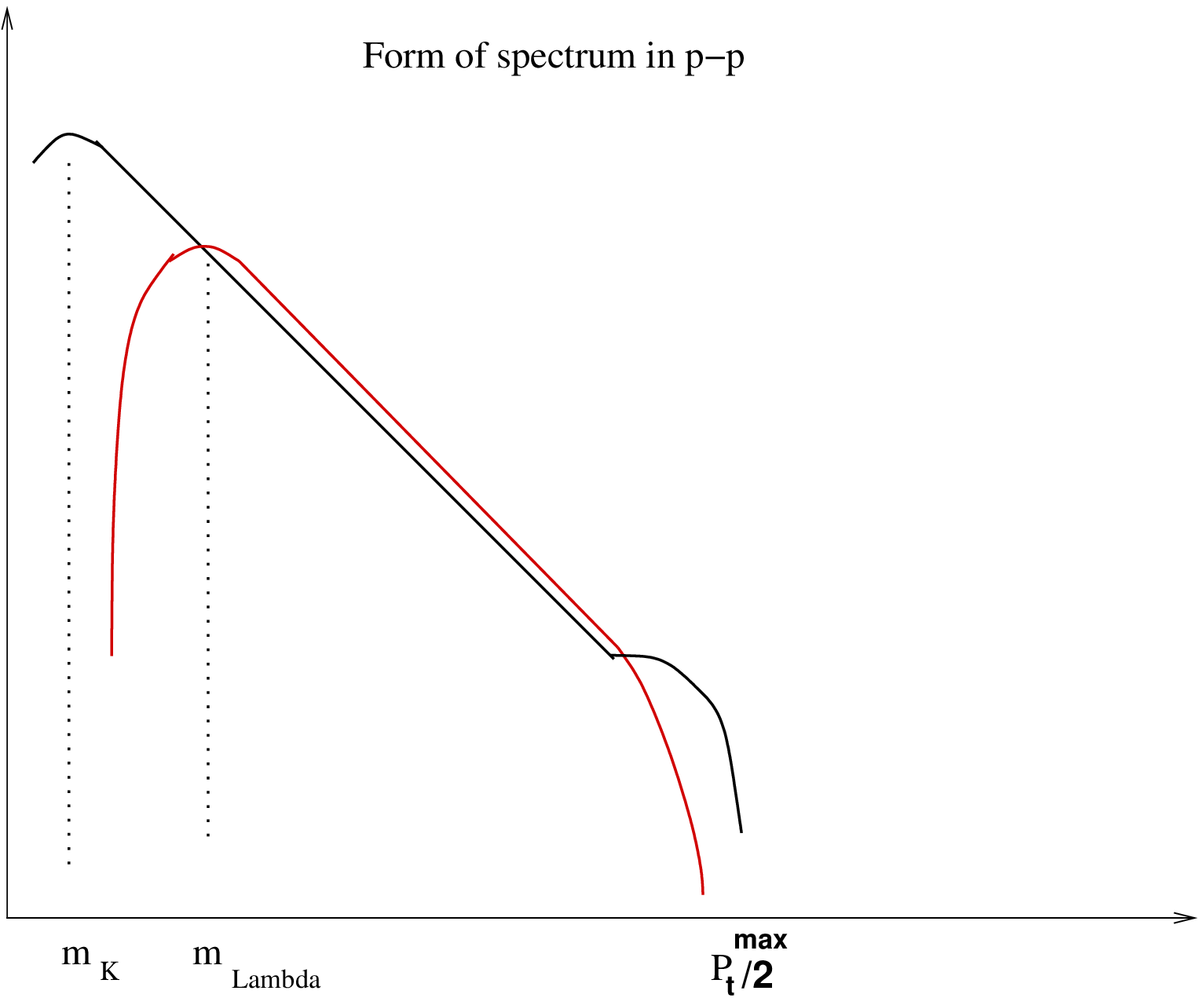,height=2.0in}
\caption{The form of spectra in proton-proton collisions.}
\label{fig:6}
\end{center}
\end{figure}

\begin{figure}[thb]
\begin{center}
\epsfig{figure=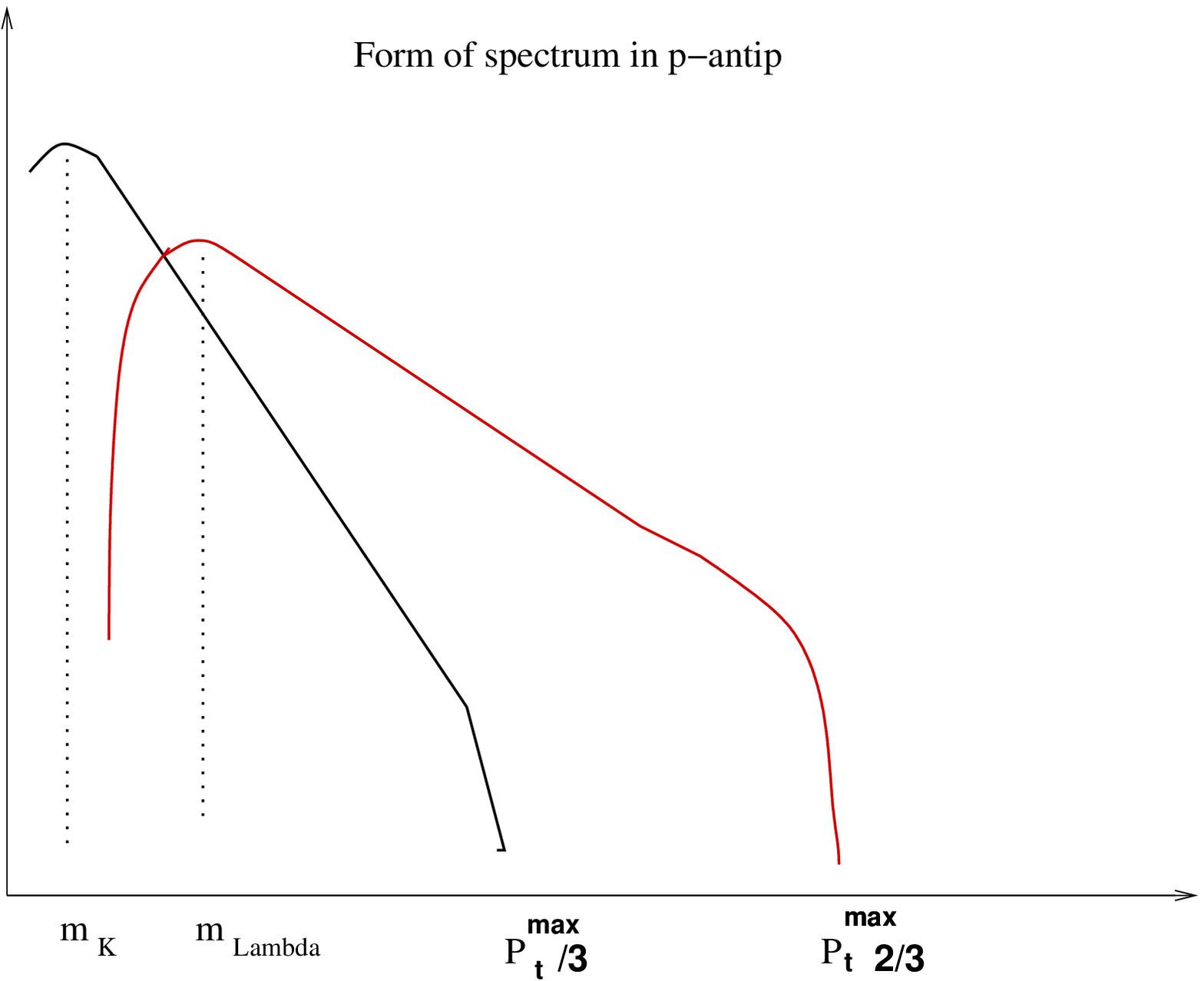,height=2.0in}
\caption{The form of spectra in antiproton-proton collisions.}
\label{fig:7}
\end{center}
\end{figure}

\end{document}